\pdfoutput=1

\documentclass[11pt]{article}

\usepackage[final]{acl}
\usepackage{amssymb}
\usepackage{times}
\usepackage{latexsym}
\usepackage{booktabs} 
 \usepackage{amsmath}
\usepackage[T1]{fontenc}
\usepackage{multirow}
\usepackage[utf8]{inputenc}
\usepackage[inline]{enumitem}
\usepackage{microtype}

\usepackage{inconsolata}

\usepackage{graphicx}

\newcommand{\heading}[1]{\vspace*{1mm}\noindent\textbf{#1.}}

\title{A Generative Framework for Personalized Sticker Retrieval}

\author{
 \textbf{Changjiang Zhou\textsuperscript{1,2}},
 \textbf{Ruqing Zhang\textsuperscript{1,2}}$^{*}$, 
 \textbf{Jiafeng Guo\textsuperscript{1,2}}$^{*}$,
 \textbf{Yu-An Liu\textsuperscript{1,2}},
\\
 \textbf{Fan Zhang\textsuperscript{3}},
 \textbf{Ganyuan Luo\textsuperscript{3}},
 \textbf{Xueqi Cheng\textsuperscript{1,2}}
\\
 \textsuperscript{1}State Key Laboratory of AI Safety, Institute of Computing Technology, CAS, Beijing, China \\
 \textsuperscript{2}University of Chinese Academy of Sciences, Beijing, China\\
 \textsuperscript{3}WeChat Search Application Department, Tencent Inc., China
\\
 \small{
   \{zhouchangjiang23s,zhangruqing,guojiafeng\}@ict.ac.cn ~\{fatumzhang,garrettluo\}@tencent.com
 }
}

\begin{document}
\maketitle
\renewcommand{\thefootnote}{\fnsymbol{footnote}} 
\footnotetext[1]{Corresponding authors}

\begin{abstract}

Formulating information retrieval as a variant of generative modeling, specifically using autoregressive models to generate relevant identifiers for a given query, has recently attracted considerable attention. 
However, its application to personalized sticker retrieval remains largely unexplored and presents unique challenges: existing relevance-based generative retrieval methods typically lack personalization, leading to a mismatch between diverse user expectations and the retrieved results. 
To address this gap, we propose PEARL, a novel generative framework for personalized sticker retrieval, and make two key contributions: 
(i) To encode user-specific sticker preferences, we design a representation learning model to learn discriminative user representations. It is trained on three prediction tasks that leverage personal information and click history; and
(ii) To generate stickers aligned with a user's query intent, we propose a novel intent-aware learning objective that prioritizes stickers associated with higher-ranked intents. 
Empirical results from both offline evaluations and online tests demonstrate that PEARL significantly outperforms state-of-the-art methods.

\end{abstract}

\section{Introduction}

With the rise of instant messaging applications, online chatting has become an integral part of daily communication.  
Stickers, as expressive visual elements commonly used on platforms such as WeChat and WhatsApp, play a crucial role in conveying emotions and sentiments. 
As users increasingly rely on stickers to express themselves, personalized sticker retrieval becomes crucial for retrieving stickers that match users' unique communication styles and emotional preferences \cite{konrad2020sticker,chee2025persrv}.

\heading{Using generative modeling for sticker retrieval} Generative retrieval (GR) is an emerging paradigm in information retrieval \cite{tay2022transformer}, where the entire corpus is encoded into model parameters, enabling a single parametric model to directly generate a ranked list of results.
Typically, a sequence-to-sequence (Seq2Seq) encoder-decoder architecture is employed to predict the identifiers of documents relevant to a given query.
Recent studies have demonstrated impressive performance across various retrieval tasks, e.g., passage retrieval and image retrieval \cite{zhang2024irgen, zhang2018image, tang2023semantic,long2024generative}.

However, directly applying existing relevant-based GR methods to personalized sticker retrieval poses unique challenges: 
\begin{enumerate*}[label=(\roman*)]
    \item \emph{Different users prefer different stickers.} Personalized sticker retrieval should incorporate user-specific information, e.g., personal portraits and historical preferences, rather than relying solely on query-sticker semantic associations as in existing GR methods. 
    For instance, given the query ``Hello'', younger users may prefer lively, animated stickers, while older users may favor more restrained or text-based ones. 
    \item \emph{A single user's preference for sticker properties varies with intent.} This calls for intent-aware ranking that aligns with the user's preferences across different sticker properties—be it character IP, visual style, or textual content. For example, for the query ``Doraemon sleeping'', sticker properties related to the Doraemon character should be prioritized. In contrast, for ``good morning'', textual content extracted via OCR may be more important.
\end{enumerate*}

\heading{A personalized sticker retriever} Our goal is to develop an effective \emph{PErsonalized-learner for generAtive sticker RetrievaL} (PEARL), that can bridge the gap between diverse user expectations and the relevant stickers retrieved by generative modeling. 
To this end, we need to resolve two key challenges in terms of encoding and decoding.

First, \emph{How to encode user-specific preferences effectively?} 
In this work, we consider that user-specific preferences are mainly determined by the user's age and gender, as well as historical click-through data.
In GR, generating document identifiers using dense document representations has been proven effective~\cite{zhou2022ultron, li2024matching}. 
However, user-specific information has not been adequately considered in existing studies. 
To address the issue, we first categorize users based on their age and gender into distinct user groups, and then for each user group, we design a discriminative representation learning model that captures the unique characteristics of the user group.
Specifically, three tasks, including user click prediction, user intent prediction and user interest prediction, are involved in the representation learning of the user group using data in the history click log:
Subsequently, the user group representation is input into the generative model along with the user query for personalized encoding.

Second, \emph{How to decode stickers that align with individual expressive intent?}
A sticker typically involves multiple properties, such as character IP, OCR textual content, visual style, entity, and meaning.
We first generate a product quantization (PQ) code for each property of a given sticker as its property identifier~\cite{zhou2022ultron}.
Accordingly, the objective of the GR model is to generate each property identifier of the corresponding stickers for a given input query.
We propose an \emph{intent-aware loss} that reweights the relevance between the input query and different property identifiers based on inferred user intent.
To infer user intent, we leverage the chain-of-thought (CoT) reasoning capabilities of large language models (LLMs) \cite{yu2023towards} to determine the intent ranking of the query with respect to each property dimension.
The intent-aware loss is designed to ensure that the property identifiers corresponding to higher-ranked intents receive greater attention.

\heading{Experiments and contributions} The effectiveness of PEARL is verified by extensive offline analyses and large-scale online tests.
PEARL significantly outperforms state-of-the-art methods, particularly in MRR@10 and Recall@10, with substantial improvements of 15\% and 18.3\%, and additionally achieves CTR improvements and GSB gains of 7.12\% and 5.98\% against the online system under the evaluation of human experts.

\section{Problem Statement}
\heading{Task description}
Given a textual input query $q$, the objective of sticker retrieval is to yield a ranked list $R$ of top-$k$ relevant stickers from a large sticker repository $\mathcal{S}=\{s_1,s_2,\dots,s_n\}$, where $n$ denotes the total number of stickers in the repository. 

As one of the most popular instant messaging platforms, WeChat is a representative application scenario of sticker retrieval \cite{zhou2017goodbye}. 
During our investigation of sticker retrieval in WeChat, five properties of stickers are considered in this work, including:
\begin{enumerate*}[label=(\roman*)]
\item \emph{OCR textual content $o$} refers to the text extracted from the sticker using Optical Character Recognition (OCR) technology. 
\item \emph{Character IP $c$} refers to Intellectual Property (IP) related to the characters depicted on the sticker, which could be a well-known character from a movie, TV show, comic book, video game, or any other form of media. 
\item \emph{Entity $e$} refers to the specific object, symbol, or concept that is primarily depicted in the sticker. 
\item \emph{Visual style $v$} refers to the specific artistic style that the sticker's design follows.
\item \emph{Meaning $m$} refers to the intended message, sentiment, or symbolism that the sticker is designed to convey, which is typically provided by the source of the sticker. 
\end{enumerate*}
A more detailed introduction of these properties is provided in \autoref{data example}.

\heading{User-specific personalization in sticker retrieval}
User-specific personalization refers to the process of retrieving stickers based on user-specific information beyond general relevance. 
Generally, the definition of user-specific personalization can vary across different sticker retrieval systems. 
In this work, based on our investigation in WeChat, we focus primarily on the personalization induced by age $a$, gender $g$, and historical interest in character IPs $H_c$ and entities $H_e$. 
We further categorize users based on age and gender, denoted as \emph{user groups}, and a user with age $a$ and gender $g$ is allocated into the user group $G_{a,g}$.

\begin{figure*}[t]
  \centering 

\includegraphics[width=\linewidth]{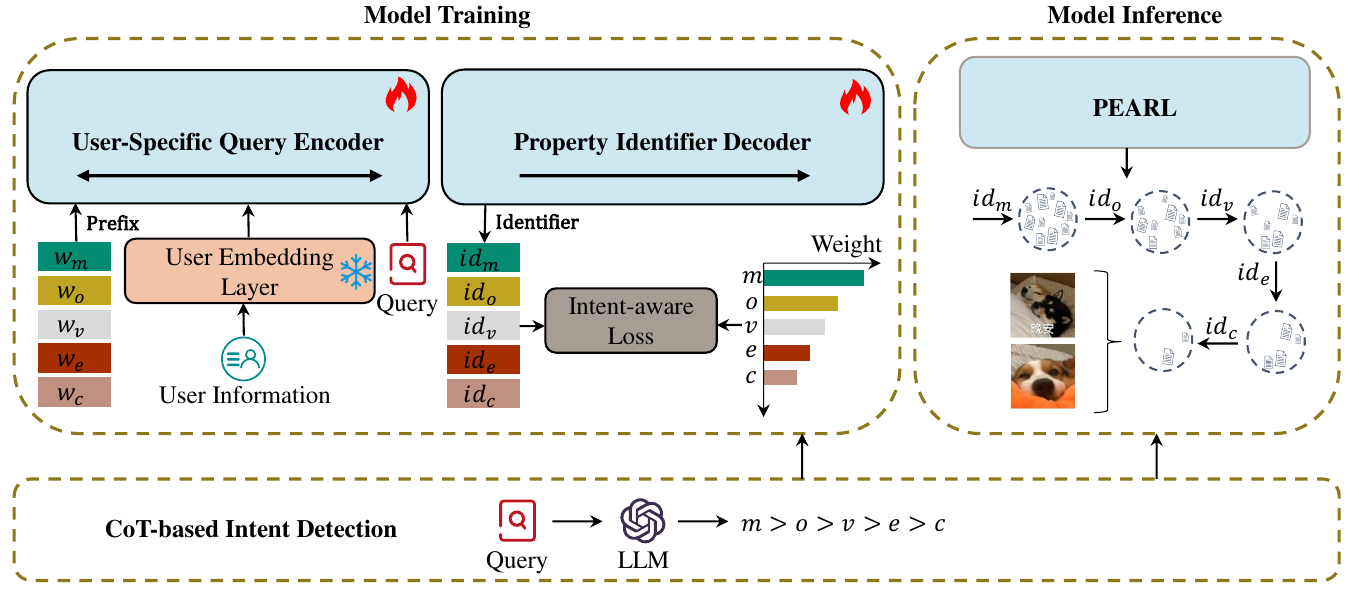}
   \vspace{-5mm} 
  \caption{The overview of PEARL.}
  \label{framework}
      \vspace{-3mm}
\end{figure*}

\heading{Benchmark construction}
In this work, we involve two sticker repositories at different scales.
\begin{enumerate*}[label=(\roman*)]
\item \emph{WeChat offline dataset}.
We construct the WeChat offline dataset by sampling partial stickers from the WeChat online system.
We enlisted human annotators for the annotation of the training and test datasets, as well as the collection of click logs with permission.
Refer to \autoref{dataset} for detailed elaboration.
\item \emph{WeChat online dataset}. We also assess retrieval performance on the online large-scale sticker repository with millions of stickers, using the internal platform of WeChat.
\end{enumerate*}

\section{Methodology}

In this section, we present the proposed PEARL for personalized sticker retrieval in detail.

\subsection{Overview}
The proposed PEARL framework employs an encoder-decoder generative  architecture: the encoder encodes the user-specific information and the query; the decoder decodes property identifiers to retrieve specific stickers.
To capture user-specific information, personalized representation learning is proposed to assign a unique dense embedding for each user group.
To align the decoding process with user intent, intent-aware loss is proposed, guiding the process of property identifier generation with user intent predicted by LLMs. 
The overview of PEARL is shown in Figure \ref{framework}.

\subsection{Model Architecture} \label{Model Architecture}
The architecture of PEARL comprises a user-specific encoder and a property identifier decoder.

\heading{User-specific query encoder}
The user-specific query encoder maps user-specific information involving the age $a$ and gender $g$ along with the input query $q=\{w_1,w_2,\dots,w_{|q|}\}$ into a compact hidden state representation, formulated as follows:
\begin{equation}
    H_q=\text{Encoder}(w_{a,g},w_1,w_2,\dots,w_{|q|}),
\end{equation}
where $H_q$ denotes the hidden state representation, and $w_{a,g}$ is a user-specific special token added to the vocabulary to represent the specific user group $G_{a,g}$ categorized by age $a$ and gender $g$.
To align the semantic representation of each user-specific token $w_{a,g}$ with user preferences, personalized representation learning is utilized to train the embedding of user-specific tokens, as presented in Section \ref{Personalized representation learning}.

\heading{Property identifier decoder}
Given the encoded representation $H_q$, the property identifier decoder is intended for yielding the property identifier of the target stickers.
Specifically, the probability of generating the $n$-th token $w_n$ in the target identifier of the property $p\in\{o,c,e,v,m\}$ is defined as:
\begin{equation}
P(w_n|w_{<n},q,a,g,p)=\text{Decoder}(w_{<n},H_q,w_p),
\end{equation}
where $w_p$ is a special token indicating the identifier start of the property $p$. The identifier construction is introduced as follows.

\heading{Sticker identifier} Since each sticker has multiple properties, we propose representing each sticker with multiple identifiers corresponding to its different properties. 
For property identifier construction, we apply semantic-based property identifiers through Product Quantization (PQ) \cite{zhou2022ultron}. 
For all $D$-dimensional vectors, PQ first partitions the $D$-dimensional space into $m$ disjoint subspaces. Subsequently, $k$-means clustering is independently applied to each subspace to obtain $k$ cluster centroids per group. Each vector is ultimately represented by a sequence of $m$ cluster identifiers, corresponding to the nearest centroids in each subspace.
More details on PQ refer to Appendix \ref{pq}. 
We leverage BERT \cite{devlin2019bert} to encode the property $p$ and then the identifier of each property for a specific sticker is defined as: 
\begin{equation}
    id_p=\text{PQ}(\text{BERT}(p)), ~~~~ p \in \{o,c,e,v,m\}, 
\end{equation}
where multiple property identifiers $id_p$ with respect to a specific sticker are treated as new tokens and added to the vocabulary.

At the inference time, the constrained beam search strategy is utilized to limit each generated identifier within a pre-defined candidate set.
The order in which different property identifiers are decoded is guided by the intent contained in the query, as in Section \ref{Intent-aware model training}.

\subsection{Personalized representation learning}\label{Personalized representation learning}
\label{Personalized representation learning}
\begin{figure}[t]
  \centering 
   \vspace{-5mm} 
  \includegraphics[width=\linewidth]{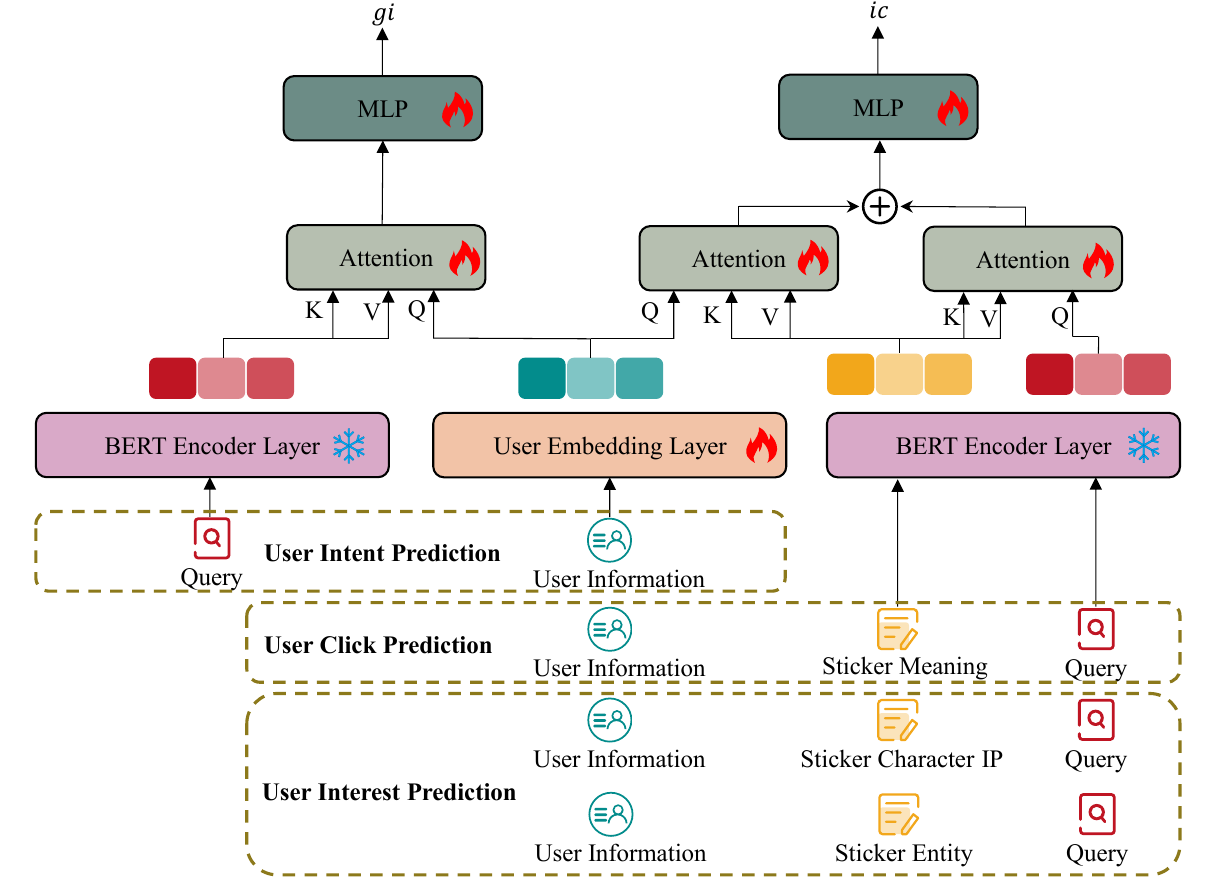}
  \caption{The learning of user-specific representation.}
  \label{user}
     \vspace{-3mm}
\end{figure}

As shown in Figure \ref{user}, we leverage additional data from user click logs for personalized representation learning, trained with three discriminative tasks.  
The training data for personalized representation learning is sampled from the user click logs dumped from the online sticker search system.
Apart from the metadata of stickers, i.e., $\{o,c,e,v,m\}$ and the user-specific information $\{a, g, H_c,H_e\}$, user logs additionally involve the input query $q$ and the click behavior $ic$ (is clicked) which indicates whether the user clicks the sticker.

For the description of three tasks, we first outline the used attention mechanism. 
Given the input hidden state $h^q, h^k, h^v \in \mathbb{R}^{d}$ , the attention mechanism $A(\textbf{h}^q,\textbf{h}^k,\textbf{h}^v)$ can be formulated as: 
\begin{equation} 
A(\cdot)=\text{softmax}\left(\frac{W^q \textbf{h}^q\cdot W^k \textbf{h}^k}{\sqrt{d}}\right)W^v\textbf{h}^v,
\end{equation}
where $W^{(\cdot)} \in \mathbb{R}^{d}$ are trainable projection matrices.

\heading{User click prediction} The core idea is to predict whether the user will click a specific sticker after sending the query.
This task directly captures the understanding of users in terms of the relevance of the query and the meaning of stickers, formulated as a binary classification task: 
\begin{equation}
h_q=A(\text{BERT}(q),\text{BERT}(m),\text{BERT}(m)),
\end{equation}
\begin{equation}
h_u=A(\text{UE}(w_{a,g}),\text{BERT}(m),\text{BERT}(m)),
\end{equation}
\begin{equation}
    \hat{ic}=\text{sigmoid}(\text{MLP}(\text{concat}(h_q,h_u))),
\end{equation}
where $\text{UE}$ denotes the user embedding layer, $\hat{}$ is utilized to notate the predicted results. Suppose the golden ``is clicked'' label is $ic$, hence the training loss for the user click prediction task is formulated as follows:
\begin{equation}
\mathcal{L}_{click}=-ic\cdot \text{log}(\hat{ic})-(1-ic)\cdot \text{log}(1-\hat{ic}).
\end{equation}

\heading{User intent prediction}
The core idea is to predict the intent preference of users hidden in the input query. 
LLMs are employed to obtain the golden intent $gi$ for a query $q$, and the prompting strategy is explained in Appendix \ref{Prompt for intent permutation generation} in detail.
This task is formulated as a multi-label classification task:
\begin{equation}
    h_{i}=A(\text{UE}(w_{a,g}),\text{BERT}(q),\text{BERT}(q)),
\end{equation}
\begin{equation}
   \hat{gi}=\text{softmax}(\text{MLP}(h_i)),
\end{equation}
where $\text{UE}$ denotes the user embedding layer, $\hat{}$ is utilized to notate the predicted results. Suppose the real ``golden intent'' label is $gi\in\{o,c,e,v,m\}$, hence the training loss for the user intent prediction task is formulated as follows:
\begin{equation}
\mathcal{L}_{intent}=-\sum_{p\in\{o,c,e,v,m\}}\mathbb{I}(gi=p)\text{log}\hat{gi},
\end{equation}
where $\mathbb{I}(.)$ denotes the indicator function.

\heading{User interest prediction}
The core idea is to predict whether a user will be interested in a specific sticker based on the user's historical click behavior. 
Distinct from the query-meaning relevance, user interest is typically influenced by the character IP and the entity in the sticker.
This task is motivated by the phenomenon that younger individuals tend to favor lively and trendy stickers, while older individuals lean towards more conservative and accessible options~\cite{konrad2020sticker}.
For the character IP interest $c$, the task can be formulated as follows:
\begin{equation}
h_q=A(\text{BERT}(q),\text{BERT}(c),\text{BERT}(c)),
\end{equation}
\begin{equation}
h_u=A(\text{UE}(w_{a,g}),\text{BERT}(c),\text{BERT}(c)),
\end{equation}
\begin{equation}
    \hat{ic}=\text{sigmoid}(\text{MLP}(\text{concat}(h_q,h_u))),
\end{equation}
where $\text{UE}$ denotes the user embedding layer,where $\hat{}$ is utilized to notate the predicted results. Suppose the golden ``is clicked'' label is $ic$, hence the training loss for the user click prediction task is formulated as follows:
\begin{equation}
\mathcal{L}_{interest}=-ic\cdot \text{log}(\hat{ic})-(1-ic)\cdot \text{log}(1-\hat{ic}).
\end{equation}
It is noteworthy that the user interest prediction task actually contains both the character IP interest and the entity interest, we omit the formalization of the entity interest in Equation. 12-14 since the process is similar for both. The aforementioned loss function applied to the entity interest as well.

\heading{Learning} The user embedding of $w_{a,g}$ is learned by jointly optimizing the aforementioned three modules with maximum likelihood estimation (MLE), and the total training loss of these user embedding learning tasks can be formulated as follows:
\begin{equation}
\mathcal{L}_{UE}=\mathcal{L}_{click}+\mathcal{L}_{intent}+\mathcal{L}_{interest}.
\end{equation}
The learned embedding of the special token $w_{a,g}$ is retained frozen for subsequent application in the generative retrieval framework.

\subsection{Intent-aware model training}\label{Intent-aware model training}
\label{Intent-aware model training}

\heading{CoT-based intent detection}
Given the input query $q$, we utilize the CoT capability of LLMs to determine the intent ranking with respect to each property dimension. Specifically, 
\begin{enumerate*}[label=(\roman*)]
\item we first prompt the LLM to perform the intent detection task by providing the introduction of different properties in $\{o,c,e,v,m\}$ with some examples. 
\item we then construct a question-answer pair that formats the LLM output: 
In the question part, we provide a specific query example.
In the answer part, we provide the reasoning process that iteratively prioritizes and explains the intent with the highest probability from the intent remaining set, discarding each selected intent until none remain.
\end{enumerate*}
A specific prompt applied in our implementation is provided in \autoref{Prompt for intent permutation generation}.

By prompting LLMs in the CoT manner, a ranked list of intended properties $\mathcal{R}$ can be yielded for each query. 
The intent detection strategy is applied to queries in both the test set and the training set, aiming to enhance the consistency between training and inference of GR models.

\heading{Model training: indexing}
The target is to memorize the information about each specific sticker. 
In this phase, the metadata within each sticker is indexed into the model parameters by mapping each property content to the property identifier, i.e.,
\begin{equation}
    \mathcal{L}_{I}=-\sum_{i=1}^{n}\sum_{p\in\{o,c,e,v,m\}}\text{log}(P_\theta(id_{p_i}|w_p,p_i)),
\end{equation}
where $n$ denotes the number of stickers in the corpus and $w_p$ is a special prefix token indicating which property identifier to generate.

\heading{Model training: retrieval}
Labeled training data involving user-query-sticker triplets
is further utilized for the integration of personalized user information.
After acquiring the ranked list of intended properties $\mathcal{R}$ for queries in the training set, we propose an intent-aware loss to reweight the relevance between the input query and different property dimensions.
The core idea is to prioritize stickers with higher-ranked intents. Suppose each user-query-sticker triplet contained in the training dataset $\mathcal{T}$ is $\tau=(G_{a,g},q,s_i)$, the optimization objective can be formulated as:
\begin{equation}
    \mathcal{L}_{R}=-\sum_{\tau\in\mathcal{T}}\sum_{p\in \mathcal{R}}d_{p}\text{log}(P_\theta(id_{p_i}|w_p,w_{a,g},q)),
\end{equation}
where $w_p$ is a special prefix token indicating which property identifier to generate. 
The decay weight $d_p$ is defined as: 
\begin{equation}
    d_p = \frac{1}{\text{log}_2(\text{rank}(p)+1)}, 
\end{equation}
where $\text{rank}(.)$ returns the intent rank within $\mathcal{R}$. 

The GR model is learned by jointly optimizing the indexing loss and the retrieval loss, and the total loss $\mathcal{L}_{T}$ can be formulated as follows:
\begin{equation}
    \mathcal{L}_T = \mathcal{L}_{I} +\mathcal{L}_{R}. 
\end{equation}

\heading{Model inference}
Given a test query $q$, the model inference phase is guided by the ranked list of intended properties $\mathcal{R}$.
\begin{enumerate*}[label=(\roman*)]
\item First, we construct an initial prefix tree for each intent, i.e., $T_o, T_c, T_e, T_v, T_m$, using property identifiers that span across all stickers.
\item When processing the $i$-th intent $p$ in the intent list $\mathcal{R}$, we perform constrained beam search during decoding on the prefix tree $T_{p}$ to obtain a series of property identifiers, which correspond to a collection of stickers $\mathcal{S}_i$.
\item We filter $\mathcal{S}_i$ by removing the stickers which do not appear in $\mathcal{S}_{i-1}$.
\item This process is iteratively repeated until all intents in $\mathcal{R}$ have been processed, resulting in the final collection of target stickers $\mathcal{S}_{|\mathcal{R}|}$.
\end{enumerate*}
With intent aware, the model inference process is performed in a funnel-like manner, transitioning from a coarse-grained to a fine-grained focus.

\section{Experimental Settings}
\heading{Implementation details} BERT corresponds to the pre-trained \texttt{bge-large-zh-v1.5}\footnote{https://huggingface.co/BAAI/bge-large-zh-v1.5}.
We adopt \texttt{bart-large}\footnote{https://huggingface.co/facebook/bart-large}
as the encoder-decoder backbone of PEARL.
We employ \texttt{deepseek-chat}\footnote{https://www.deepseek.com/} for CoT-based intent detection.
For PQ, the number of subspaces $m$ is 8, and the number of clusters $k$ is  256.
During inference, we set the beam size to 10 and maximum decoding steps to 15. 
Refer to \autoref{detail} for more implementation details.
 
\begin{table*}[t]
\resizebox{\textwidth}{!}{
\renewcommand{\arraystretch}{1.12}
\begin{tabular}{llcccccccc}
\hline
\multirow{2}{*}{}                         & \multirow{2}{*}{\textbf{Model}} & \multicolumn{4}{c}{\textbf{MRR}}                                                       & \multicolumn{4}{c}{\textbf{Recall}}                                                                                 \\ \cmidrule(r){3-6} \cmidrule(r){7-7} \cmidrule(r){7-10} 
                                          &                                 & \textbf{@1}                   & \textbf{@5}                & \textbf{@10}                  & \textbf{@20}                  & \textbf{@1}                   & \textbf{@5}                                & \textbf{@10}                  & \textbf{@20}                               \\ \hline
\multirow{2}{*}{\emph{Popularity-based}} & GPop                            & 0.0029               & 0.0069            & 0.0069               & 0.0069               & 0.0002               & 0.0012                            & 0.0012               & 0.0012                            \\ 
                                          & UPop                            & 0.0231               & 0.0308            & 0.0315               & 0.0319               & 0.0024               & 0.0055                            & 0.0061               & 0.0067                            \\ \hline
\multirow{3}{*}{\emph{Traditional}}    & BM25                            & 0.0519               & 0.0719            & 0.0783               & 0.0826               & 0.0049               & 0.0195                            & 0.0282               & 0.0486                            \\ 
                                          & DPR                             & 0.0778               & 0.1175            & 0.1314               & 0.1385               & 0.0087               & 0.0256                            & 0.0486               & 0.0705                            \\ 
                                          & ANCE                            & 0.0823               & 0.1293            & 0.1454               & 0.1478               & 0.0172               & 0.0328                            & 0.0592               & 0.0793                            \\ \hline
\multirow{3}{*}{\emph{Cross-modal}}   & CN-CLIP                         & 0.0375               & 0.0780             & 0.0798               & 0.0800                  & 0.0046               & 0.0198                            & 0.0223               & 0.0228                            \\ 
                                          & StickerCLIP                     & 0.0528               & 0.0821            & 0.0842               & 0.0892               & 0.0052               & 0.0203                            & 0.0235               & 0.0248                            \\ 
&PerSRV &0.1061 &0.1328 &0.1401 &0.1496 &0.0129 &0.0318 &0.0476 &0.0617          \\                               
                                          
                                          \hline
\multirow{5}{*}{\emph{Generative}}    & DSI                             & 0.0029               & 0.0079            & 0.0079               & 0.0079               & 0.0002               & 0.0010 & 0.0011               & 0.0010 \\ 
                                          & DSI-QG                          & 0.0000               & 0.0033            & 0.0048               & 0.0062               & 0.0000                      & 0.0018                            & 0.0028               & 0.0084                            \\ 
                                          & GENRE                           & 0.0317               & 0.0512            & 0.0539               & 0.0543               & 0.0039               & 0.0104                            & 0.0143               & 0.0152                            \\ 
                                          & MINDER                          & 0.1327               & 0.1699            & 0.1804               & 0.1987               & 0.0167               & 0.0492                            & 0.0594               & 0.0703                            \\ 
                                          & PEARL                           & \textbf{\phantom{0}0.1547}$^*$               & \textbf{\phantom{0}0.1839}$^*$            & \textbf{\phantom{0}0.2074}$^*$               & \textbf{\phantom{0}0.2143}$^*$               & \textbf{\phantom{0}0.0288}$^*$               & \textbf{\phantom{0}0.0582}$^*$                            & \textbf{\phantom{0}0.0732}$^*$               & \textbf{\phantom{0}0.0835}$^*$                            \\ \hline
                                          
\end{tabular}
}
\vspace{-1mm}
\caption{Retrieval performance of PEARL and the baselines on the WeChat offline dataset. $*$ indicates statistically significant improvements over the best performing baseline MINDER ($p\le0.05$).} \label{main}
\vspace{-1mm}
\end{table*}

\heading{Evaluation metrics}
We adopt two evaluation metrics: 
\begin{enumerate*}[label=(\roman*)]
\item \emph{Mean reciprocal rank (MRR@k)} measures the relative ranking position of positive stickers. We use MRR@\{1,5,10,20\} in our settings.
\item \emph{Recall@k} measures whether positive stickers are ranked in the top-k candidate list. We use Recall@\{1,5,10,20\} in our settings.
\end{enumerate*}

\heading{Baseline methods}
We compare PEARL's retrieval effectiveness with four categories of representative methods:
\begin{enumerate*}[label=(\roman*)]
\item \emph{Popularity-based methods}: Global Popularity (GPop) that returns the most popular stickers globally and User Group Popularity (UPop) that independently returns the most popular stickers for each user group.
The popularity is obtained from the online click log statistics of the WeChat system.
\item \emph{Traditional retrievers}: BM25~\cite{steck2011item}, DPR~\cite{karpukhin2020dense} and ANCE~\cite{xiong2020approximate}.
\item \emph{Cross-modal retrievers}: CN-CLIP~\cite{yang2022chinese}, StickerCLIP~\cite{zhao2023sticker820k}, and PerSRV~\cite{chee2025persrv}.
\item \emph{Generative retrievers}: DSI~\cite{tay2022transformer}, DSI-QG~\cite{zhuang2022bridging}, GENRE~\cite{de2020autoregressive},  Ultron~\cite{zhou2022ultron} and MINDER~\cite{li2023multiview}.
\end{enumerate*}

\heading{Model variants}
To validate the effectiveness of each components in PEARL, we implement the following variants to facilitate ablation studies:
\begin{enumerate*}[label=(\roman*)]
\item PEARL$_{-UE}$ removes the user embedding from the framework and ignores variability in queries from different user groups.
\item PEARL$_{click}$ only employs the task of user click prediction in Section \ref{Personalized representation learning} to train the user embedding.
\item PEARL$_{intent}$ only employs the task of user intent prediction in Section \ref{Personalized representation learning} to train the user embedding.
\item PEARL$_{interest}$ only employs the task of user interest prediction in Section \ref{Personalized representation learning} to train the user embedding.
\item PEARL$_{-IAL}$ removes the intent-aware loss in Section \ref{Intent-aware model training} during the model training phase.
\item PEARL$_{-IG}$ removes the intent-guided docid decoding process in Section \ref{Intent-aware model training} during the model inference phase and considers the intent of the user query to be equivalent.
\end{enumerate*}

\section{Experimental Results}
\subsection{Main results}
\autoref{main} shows the comparison of PEARL and baselines on the WeChat dataset.

\heading{Popularity-based methods}
We find that: 
\begin{enumerate*}[label=(\roman*)]
\item UPop, which independently returns the most popular stickers for each user group, exhibits superior retrieval capability than GPop, which neglects the differences between user groups. The phenomenon highlights the importance of preference differences among different user groups.
\item PEARL significantly outperforms popularity-based methods. The underlying reason is that popularity-based methods focus exclusively on the popularity of stickers while neglecting the relevance between queries and stickers.
\end{enumerate*}

\heading{Traditional retrievers}
When it comes to traditional retrievers including BM25, DPR and ANCE, PEARL outperforms all traditional retrievers in terms of retrieval performance. The underlying reason might be that PEARL models user preferences into generative models instead of simply relying on relevance between queries and stickers. 

\heading{Cross-modal retrievers}
We can conclude as follows:
\begin{enumerate*}[label=(\roman*)]
    \item Although a new image modality is introduced, cross-modal retrievers do not demonstrate the anticipated improvement in retrieval performance. In fact, the performance of cross-modal retrievers lags behind that of text-based dense retrievers.
    The underlying reason might be that the image modality of stickers tends to be diverse and expressive, hence posing significant challenges and difficulties for modal alignment.
    \item PEARL and PerSRV both model user preference for stickers, and PEARL exhibits superior retrieval performance. We attribute the phenomenon to the fact that apart from modeling user preference for stickers, PEARL further mines user intent behind queries, leading to more specific personalization.
\end{enumerate*}

\heading{Generative retrievers}
When we look at generative retrievers, we can find that:
\begin{enumerate*}[label=(\roman*)]
\item Approaches applying multi-view docids, including MINDER and PEARL, significantly outperforms other methods utilizing either naive string docids (DSI and DSI-QG) or meaning-based single-view docids (GENRE).
\item PEARL outperforms all other generative baselines. The underlying reason might be that the personalized representation learning and the intent-aware model training are devised tailor for personalized sticker retrieval.
\end{enumerate*}

\begin{table}[t] 
\begin{tabular*}{\linewidth}{@{\extracolsep{\fill}}lcc}
\toprule
\textbf{Model} & \textbf{MRR@10} & \textbf{Recall@10} \\ 
\midrule
PEARL &0.2074 &0.0732  \\ \midrule
\multicolumn{3}{l}{\emph{w/o personalized user embedding}} \\
PEARL$_{-UE}$ &0.1497 &0.0463 \\ 
PEARL$_{click}$ &0.1639 &0.0585 \\
PEARL$_{intent}$ &0.1563 &0.0518 \\
PEARL$_{interest}$ &0.1838 &0.0614 \\
\midrule
\multicolumn{3}{l}{\emph{w/o intent-aware loss}} \\
PEARL$_{-IAL}$ &0.1863 &0.0638\\ \midrule
\multicolumn{3}{l}{\emph{w/o intent guidance}}\\
PEARL$_{-IG}$ &0.1782 &0.0575 \\
 \bottomrule
\end{tabular*}
\vspace{-2mm}
\caption{Ablation study on the WeChat offline dataset.} \label{ablation}
\vspace{-2mm}
\end{table}

\subsection{Ablation studies}
To further validate the effectiveness of each module in PEARL, we conduct ablation studies and report the retrieval performance of model variants in \autoref{ablation}.
The following conclusions can be drawn:
\begin{enumerate*}[label=(\roman*)]
\item The proposed personalized user embedding demonstrates the most significant contribution to retrieval effectiveness, followed by intent guidance during the inference phase, and subsequently by the incorporation of intent-aware loss during the training phase.
This highlights that sticker retrieval is an expressive and fuzzy retrieval task which relies on not only the relevance relationship between queries and stickers but also the user preference.
\item The user interest prediction task contributes most to personalized representation learning. This phenomenon illustrates that user preference for stickers primarily focuses on the preference for Character IPs and entities.
\end{enumerate*}
Moreover, we also explore the impact of distinct categories of property identifiers on the retrieval performance.
Refer to \autoref{analysis_identifier} for more details.

\subsection{Efficiency analysis}
We compare the efficiency of DPR, MINDER, and PEARL.
Note that the intent list of queries is precomputed in PEARL. 
Refer to \autoref{detail} for more details.
As depicted in \autoref{efficiency}, 
\begin{enumerate*}[label=(\roman*)]
\item Generative retrievers, i.e., MINDER and PEARL, have a significant reduction of memory footprint and inference time compared to the dense retrieval model DPR.
The reduction of memory footprint primarily lies in the elimination of the explicit document index, and the inference time decreases since the heavy retrieval process over the large-scale dense index is replaced with a light generative process over the prefix tree.
\item Compared to MINDER, PEARL requires longer inference time due to the addition of the intent-aware funnel-like decoding process.
However, we believe that such an efficiency sacrifice is worthwhile, as PEARL achieves significant effectiveness gains compared to MINDER according to \autoref{main}.
\end{enumerate*}

\begin{table}[t] 
\begin{tabular}{lccc} 
\toprule
\textbf{Model} & \textbf{Memory} & \textbf{Parameters} & \textbf{Time} \\ 
\midrule
DPR &3.6G &110M &179ms \\
MINDER &1.6G &406M &112ms \\
PEARL &1.6G &406M &124ms \\
 \bottomrule
\end{tabular}
\vspace{-2mm}
\caption{Comparisons on the memory, the number of model parameters and inference time per query.} 
\label{efficiency}
\vspace{-2mm}
\end{table}

\subsection{Online tests}
User preferences for stickers are highly subjective, hence the annotation of the golden truth data is usually incomplete in the sticker retrieval task.
To this end, we conduct an online test to further verify the effectiveness of our method. 
It is worth noting that, due to privacy issues, the online WeChat system we compare is a variant that turns off personalization at the individual user's granularity.

\heading{Evaluation} 
We compare PEARL to online WeChat systems at both the sticker and the session level for a more holistic and fair assessment.

For the sticker-level assessment, we assess PEARL and online systems with the Balanced Interleaving (BI) process~\cite{schuth2015predicting}.
The specific procedures are as follows:
\begin{enumerate*}[label=(\roman*)]
\item At the start of each query session, a fair Bernoulli trial decides which system—PEARL or the online system—drafted the first sticker.
\item The active drafter appended its next unseen sticker to the interleaved list, after which drafting control immediately passed to the other system.
\item Drafting continued in strict alternation until both original top-10 lists were exhausted, resulting in a 20-item interleaved ranking.
\item Every position in the final list was annotated with a binary ownership label, thereby enabling later attribution of each user click to its originating system.
\end{enumerate*}
The procedure preserved each model’s internal order, and the ownership of returned stickers is completely blind to users to ensure the fairness of comparison.
Twenty human experts of different ages and genders are chosen to enter queries and perform clicking behavior, leading to 1,000 valid queries.
The evaluation metrics in the sticker-level assessment are two-fold: $\Delta\mathrm{CTR}$ and $\Delta\mathrm{ACP}$, refer to \autoref{online} for a detailed introduction of the metrics.

For the session-level assessment, we show the exposure session returned by PEARL and the online system containing the top-10 stickers, without allowing the user to know which model the exposure page was derived from.
We subsequently ask the users to make an overall assessment of the preference for the exposure sessions, which is limited to three responses: \textit{preferring the left exposure session}, \textit{preferring the right exposure session}, and \textit{preferring both equally}.
Here, we measure the relative gain with $\Delta\mathrm{GSB}$, refer to \autoref{online} for a detailed introduction of the metric.
Twenty human experts of different ages and genders are chosen to enter queries and assess preference for exposure sessions, resulting in 1,000 valid queries.

\begin{table}[t] 
\centering
\begin{tabular}{ccc} 
\toprule
\textbf{$\Delta\mathrm{CTR}\uparrow$} & \textbf{$\Delta\mathrm{ACP}\downarrow$} & \textbf{$\Delta\mathrm{GSB}\uparrow$} \\ \midrule
+7.12\% &-0.19 &+5.98\% \\
 \bottomrule
\end{tabular}
\vspace{2mm}
\caption{Comparison with the online WeChat system.}\label{online test} 
\vspace{-2mm}
\end{table}

\heading{Experimental results}
As depicted in \autoref{online test}, compared to the results returned by the online system, PEARL increases the click-through-rate by 7.12\% and decrease the average-click-position by 0.19 in the sticker-level human expert evaluation.
Furthermore, we can also find that PEARL has achieved significant positive gains in terms of session-level assessment.

\begin{figure}[t]
  \centering \includegraphics[width=\linewidth]{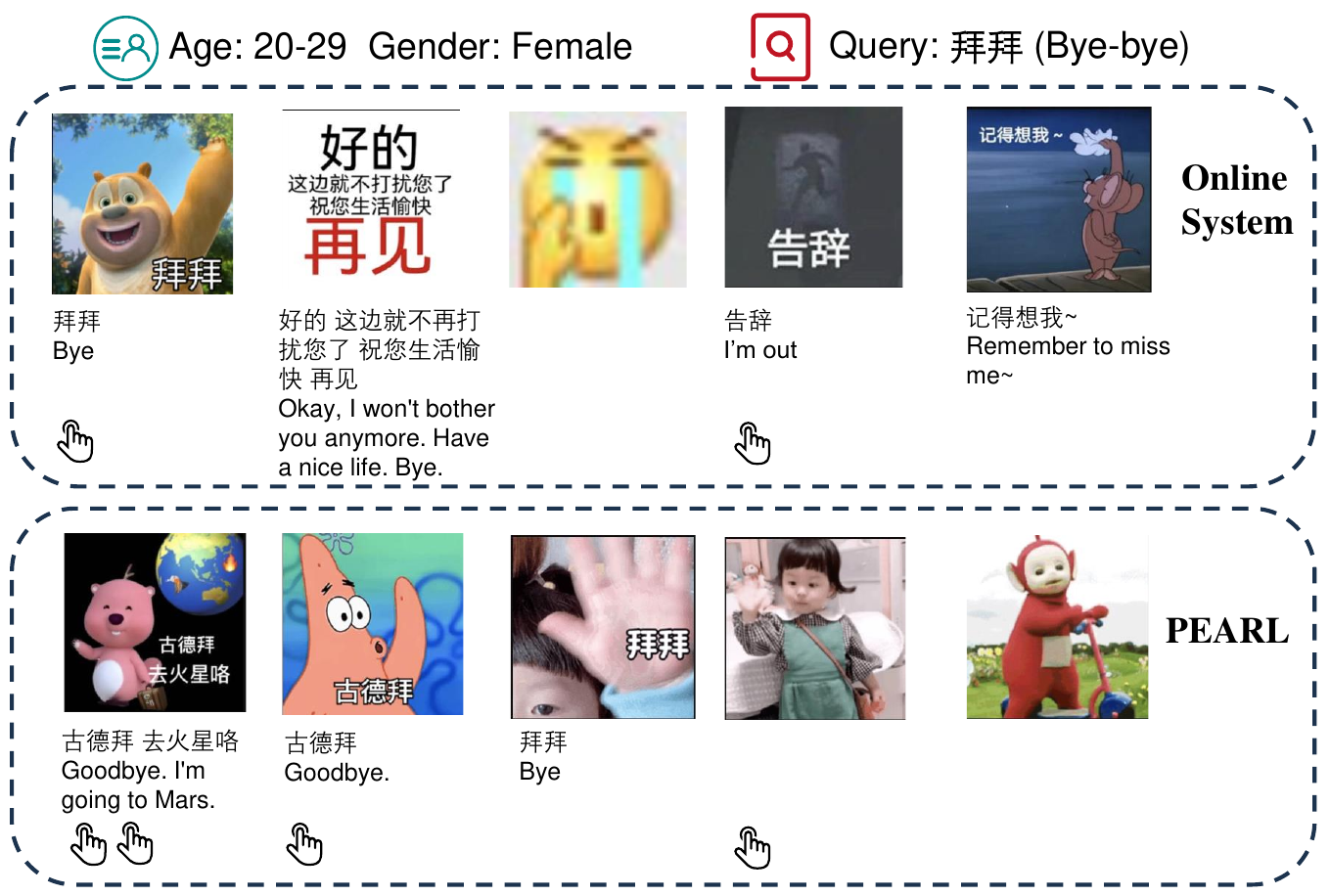}
  \vspace{-2mm} 
  \caption{Case study on retrieved results of online system and PEARL.}
  \label{case}
     \vspace{-2mm}
\end{figure}

\heading{Case study}
\autoref{case} shows the list of the top-5 stickers returned by the online system and PEARL, and the statistics of these users' clicking behavior. Our method returns stickers that are more clicked for the user query ``Bye-bye'' by female users aged 20–30. 
More cases refer to \autoref{more case}.

\section{Related work}
\heading{Sticker retrieval}
Stickers have gained significant popularity due to their ability to convey emotions, reactions, and nuanced intentions that are difficult to express through plain text~\cite{zhao2023sticker820k}.
To retrieve satisfactory stickers for users, \citet{liang2024reply} proposed a framework dubbed Int-RA based on the learning of intention and the cross-modal relationships between conversation context and stickers.  
\citet{zhao2023sticker820k} first adapted the CLIP~\cite{radford2021learning} model tailored for the domain of emotive stickers.
Most recently, PerSRV~\cite{chee2025persrv} first focused on personalized sticker retrieval and introduced user preference modeling by style-based personalized ranking.
Despite previous efforts, personalized sticker retrieval has not benefited from generative models, which have triggered transformative shifts in various areas.

\heading{Generative retrieval}
Different from the conventional methods that typically formulate information retrieval as a discriminative task~\cite{steck2011item, karpukhin2020dense,guo2019matchzoo,xia2015learning}, generative retrieval (GR) is a new retrieval paradigm in which a single consolidated model is employed to enable the direct generation of relevant docids from queries~\cite{li2024matching}.
To achieve this, two primary procedures are involved
~\cite{tay2022transformer,chen2022corpusbrain,bevilacqua2022autoregressive}, i.e., the indexing process and the retrieval process.
The indexing process learns the relationship between documents and the corresponding docids.
The retrieval process maps queries to relevant docids.
To model personalized user preference in generative retrieval, \citet{wu2024hi} proposed an efficient hierarchical encoding-decoding generative retrieval method for large-scale personalized E-commerce search systems.
Distinct from personalized E-commerce search, which typically involves specific items, the task of personalized sticker retrieval primarily focuses on the abstract expressive intent of stickers and user preference for Character IP and sticker style.
The fundamental characteristics of stickers highlight that 
Personalized generative retrieval tailored for stickers is a non-trivial challenge worth exploring.

\section{Conclusion}
In this paper, we focus on personalized sticker retrieval with the promising generative retrieval paradigm.
Since the sticker retrieval task highly calls for user personalization beyond relance relationships, we propose PEARL, a novel generative framework with user-specific information encoding and intent-aware sticker decoding.
Empirical results from both offline evaluations and online experiments indicate the superiority of PEARL.

\section*{Limitations}
The limitations of this work can be concluded as follows:
\begin{enumerate*}[label=(\roman*)]
    \item Given the importance of individual privacy, our focus is primarily on personalization at the level of user groups. This approach, however, offers a relatively coarse granularity that does not allow for the customization of sticker search and recommendations based on each individual's specific sticker preferences.
    \item For search efficiency considerations, we model only the textual information in PEARL without modeling the information of image modality. The introduction of image modality has the potential to further enhance the retrieval.
    \item The generative framework PEARL is coupled to the scenario of sticker retrieval, hence leading to restricted method generalizability.
    \item The application of LLMs for intent detection increases economic costs, restricting the large-scale industry applications. 
\end{enumerate*}

\section*{Ethical Considerations}
In this paper, all the models used in our experiment are publicly released.
For datasets, we construct offline datasets based on the open-source dataset and extra manual annotation. 
We invite human annotators for manual annotation and pay the annotators a salary that is in line with the local pay scale. 
In this process, user privacy is protected, and no personal information is contained in the dataset.
Additionally, the methods we propose aim to enhance the effectiveness and personalization of sticker retrieval and do not encourage or induce the model to produce any harmful information or leakage of user privacy.
Therefore, we believe that our research work meets the ethics of ACL.

\section*{Acknowledgments}
This work was funded by the Strategic Priority Research Program of the CAS under Grants No. XDB0680102, the National Natural Science Foundation of China (NSFC) under Grants No. 62472408 and 62441229, the National Key Research and Development Program of China under Grants No. 2023YFA1011602. 
All content represents the opinion of the authors, which is not necessarily shared or endorsed by their respective employers and/or sponsors.

\clearpage
\bibliography{main}

\clearpage
\appendix 
\label{sec:appendix}
\section*{Appendix}

\begin{figure*}[t]
  \centering 
  \includegraphics[width=\linewidth]{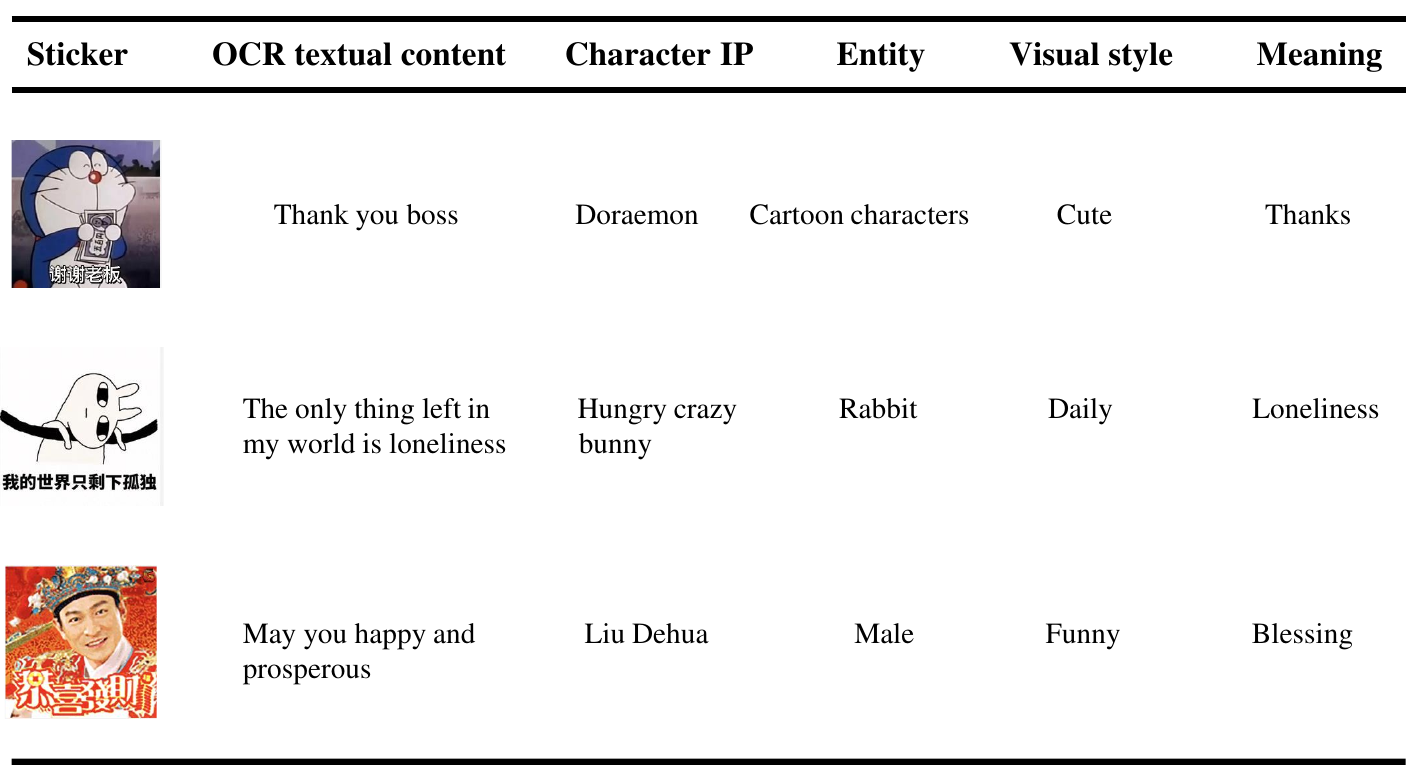}
  \caption{Examples for distinct properties of stickers in the corpus.}
  \label{example}
\end{figure*}

\section{WeChat offline dataset}\label{dataset}
We constructed a sticker corpus by sampling about 1.1 million stickers from the WeChat online system.
Offensive, potentially harmful, and copyright-controversial stickers were filtered out.
Specifically, the sticker corpus contains 1,092,122 stickers spanning 17,906 Character IPs, 38,895 entities, and 107 visual styles. 
Based on the actual usage of the sticker search function, we categorized users into four age groups (0-19,20-29,30-44, and 45-59) and two gender groups (male and female).
We enlisted human annotators across all these user groups.
We collect the user click logs with their permission and invite them to perform data annotation for both the training and test datasets.
Specifically, the training dataset contains 1,891 unique queries, 2,308 user-query pairs, and 12,568 user-query-sticker triplets.
The test dataset contains 258 unique queries, 347 user-query pairs, and 14,446 user-query-sticker triplets.
The full text of the instructions for annotating the training and the test datasets given to participants is as follows:
\texttt{Determine whether a given query and sticker match based on your personal preferences by selecting either ``Match'' or ``No Match''. The data collected will only be used to carry out research to improve the effectiveness of sticker retrieval.
In this process, user privacy is protected, and no personal information is contained in the dataset.}

We invited human annotators from the crowdsourcing platform and paid the annotators a salary that is in line with the local pay scale.
Due to the limited community of WeChat software users, we enlisted all data annotators from China.
The data collection protocol was approved by an ethics review board.
We manually filtered all collected data to remove any user privacy information.
All data used contains neither information that uniquely identifies individual people nor offensive content.

\section{Prompt for intent permutation generation}\label{Prompt for intent permutation generation}
The prompt applied in our implementation is as follows:\\
\texttt{
I am a user who is using the sticker search feature, and I have entered a query. Please help me analyze the intent behind my query.\\
There are five possible intents: OCR, IP, entity, style, and meaning.
Here are the descriptions and examples for each intent.\\
OCR textual content refers to the text extracted from the sticker using Optical Character Recognition (OCR) technology. \\
Examples: \{examples for the OCR intent\} \\
Character IP refers to Intellectual Property (IP) related to the characters depicted on the sticker, which could be a well-known character from a movie, TV show, comic book, video game, or any other form of media.\\
Examples: \{examples for the IP intent\} \\
Entity refers to the specific object, symbol, or concept that is primarily depicted in the sticker. \\
Examples: \{examples for the entity intent\} \\
Visual style refers to the specific artistic style that the sticker's design follows.\\
Examples: \{examples for the style intent\} \\
Meaning refers to the intended message, sentiment, or symbolism that the sticker is designed to convey, which is typically provided by the source of the sticker. \\
Examples: \{examples for the meaning intent\} \\
Q: Based on the given explanation, arrange the order of intent for the query: Doraemon cute.\\
A: Let's think step by step. "Doraemon cute" is most likely to be an IP intent in OCR, IP, entity, style, meaning, because Doraemon is a well-known anime character. Excluding the IP intent, among the remaining OCR, entity, style, meaning, "Doraemon cute" is most likely to be a style intent, because the query includes the style description "cute". Excluding IP and style intents, among the remaining OCR, entity, meaning, "Doraemon cute" is most likely to be an entity intent, because Doraemon is a specific character. Excluding IP, style, and entity intents, among the remaining OCR and meaning, "Doraemon cute" is most likely to be a meaning intent, because "Doraemon cute" can be understood as a certain meaning. "Doraemon cute" is least likely to be an OCR intent, because it is not an image or video with text content. Therefore, the answer is: IP > style > entity > meaning > OCR.\\
Q: Based on the given explanation, arrange the order of intent for the query: \{query\} \\
A: Let's think step by step.
}

\section{Product quantization}\label{pq}
Product quantization (PQ) is an efficient technique for approximate nearest neighbor (ANN) search in high-dimensional spaces, commonly used in large-scale retrieval tasks. It works by decomposing a $D$-dimensional vector space into $m$ low-dimensional subspaces, i.e., each input vector $\mathbf{x} \in \mathbb{R}^D$ is split into $m$ sub-vectors $\mathbf{x} = [\mathbf{x}^1, \mathbf{x}^2, \dots, \mathbf{x}^m]$, where each $\mathbf{x}^i \in \mathbb{R}^{D/m}$. In each subspace, a separate codebook is learned via $k$-means clustering, and sub-vectors are quantized by mapping them to their nearest centroids. The full vector is then represented as a concatenation of centroid indices, significantly reducing storage requirements. During search, the distance between a query vector and database vectors is approximated efficiently using precomputed lookup tables, enabling fast and memory-efficient similarity computation without reconstructing full vectors.

\section{Sticker properties} \label{data example}
For each sticker, five properties are annotated—OCR textual content, character IP, entity, visual style, and meaning. As for these five properties, the annotation methods are as follows: (i) The OCR textual content is derived by applying Optical Character Recognition (OCR) tools to each sticker. (ii) The character IP is obtained by applying a vision-language pre-trained model. (iii) The remaining properties, i.e., the meaning, the entity,  and the visual style, are primarily obtained from the tags provided by the sticker creators or the original sources.  For cases where tags are missing, a vision-language pre-trained model is employed to supplement and complete these tags.
Detailed examples of the properties in the sticker corpus are provided in \autoref{example}.

\section{Online evaluation metrics} \label{online}
For the sticker-level assessment, we report the relative advantage of PEARL over the baseline with two per-query paired-difference metrics: $\Delta\mathrm{CTR}$ and $\Delta\mathrm{ACP}$.

\heading{Click-through-rate difference}
For each query $q$, let $\mathrm{CTR}_{P}(q)$ and $\mathrm{CTR}_{B}(q)$ denote the fractions of exposed stickers that were clicked for PEARL and the baseline, respectively.
The evaluation metric $\Delta\mathrm{CTR}$ is defined as 
\begin{equation}
    \Delta\mathrm{CTR}=\frac{1}{|\mathcal{Q}|}\sum_{q\in \mathcal{Q}}(\mathrm{CTR}_{P}(q)-\mathrm{CTR}_{B}(q)),
\end{equation}
where $\mathcal{Q}$ denotes the collections of all queries.

\heading{Average-click-position difference}
Let $\mathrm{ACP}_{P}(q)$ and $\mathrm{ACP}_{B}(q)$ be the mean rank positions of the clicks attributed to each system.
The evaluation metric $\Delta\mathrm{ACP}$ is defined as 
\begin{equation}
    \Delta\mathrm{ACP}=\frac{1}{|\mathcal{Q}|}\sum_{q\in \mathcal{Q}}(\mathrm{ACP}_{P}(q)-\mathrm{ACP}_{B}(q)),
\end{equation}
where $\mathcal{Q}$ denotes the collections of all queries.
A negative value indicates that PEARL receives clicks closer to the top of the interleaved list.

For the session-level assessment, we report the relative gain of PEARL over the baseline with the metric $\Delta\mathrm {GSB}$, which can be defined as follows:
\begin{equation}
    \Delta\mathrm {GSB}=\frac{\#Good-\#Bad}{\#Good+\#Same+\#Bad},
\end{equation}
where $\#Good$ (or $\#Bad$) indicates the number of queries that the PEARL provides better (or worse) final results against the baseline.

\begin{figure}[t]
  \centering \includegraphics[width=\linewidth]{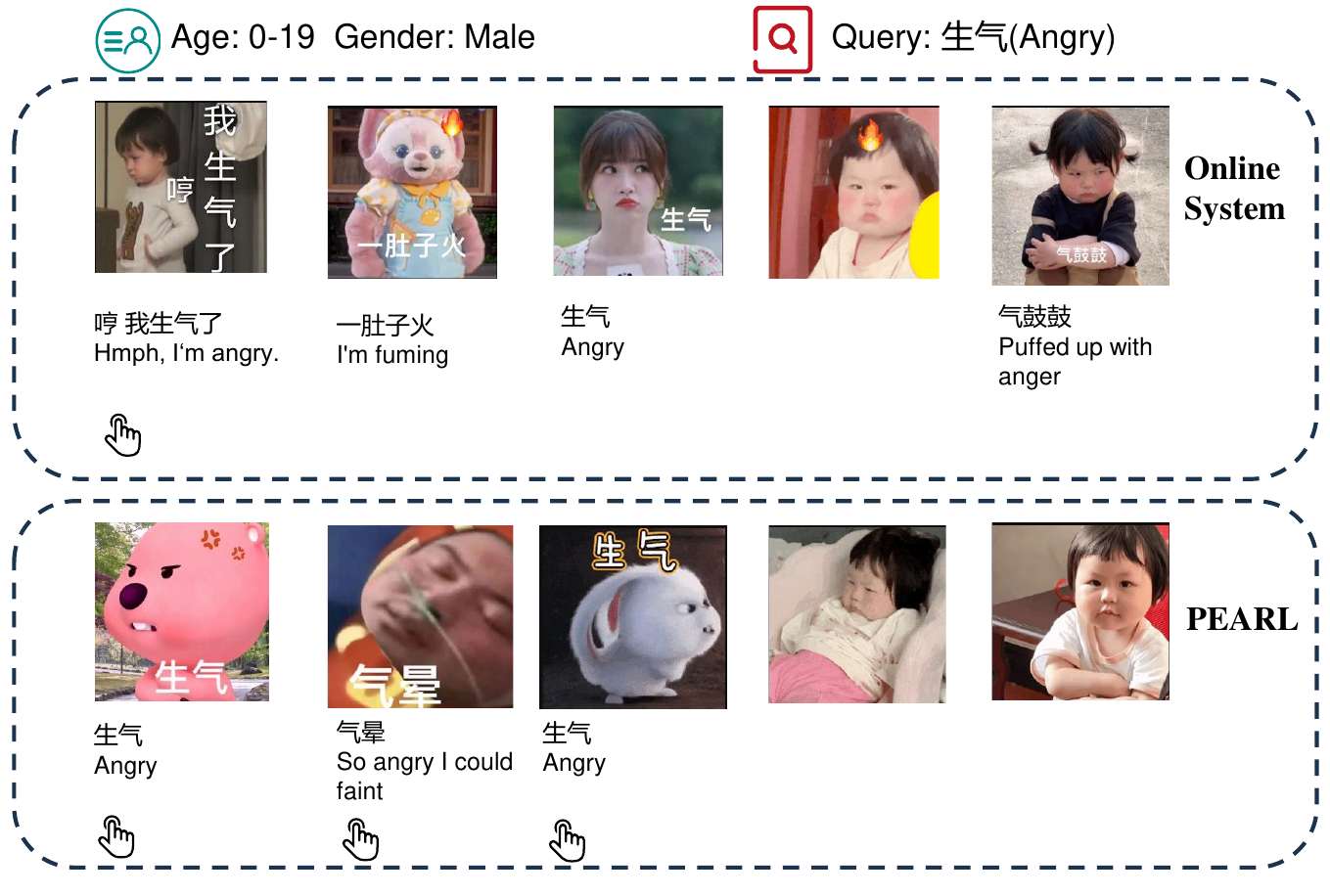}
  \vspace{-6mm} 
  \caption{Case study for the user query ``Angry'' by male users aged 0-19.}
  \label{extra1}
     \vspace{-3mm}
\end{figure}

\begin{figure}[t]
  \centering \includegraphics[width=\linewidth]{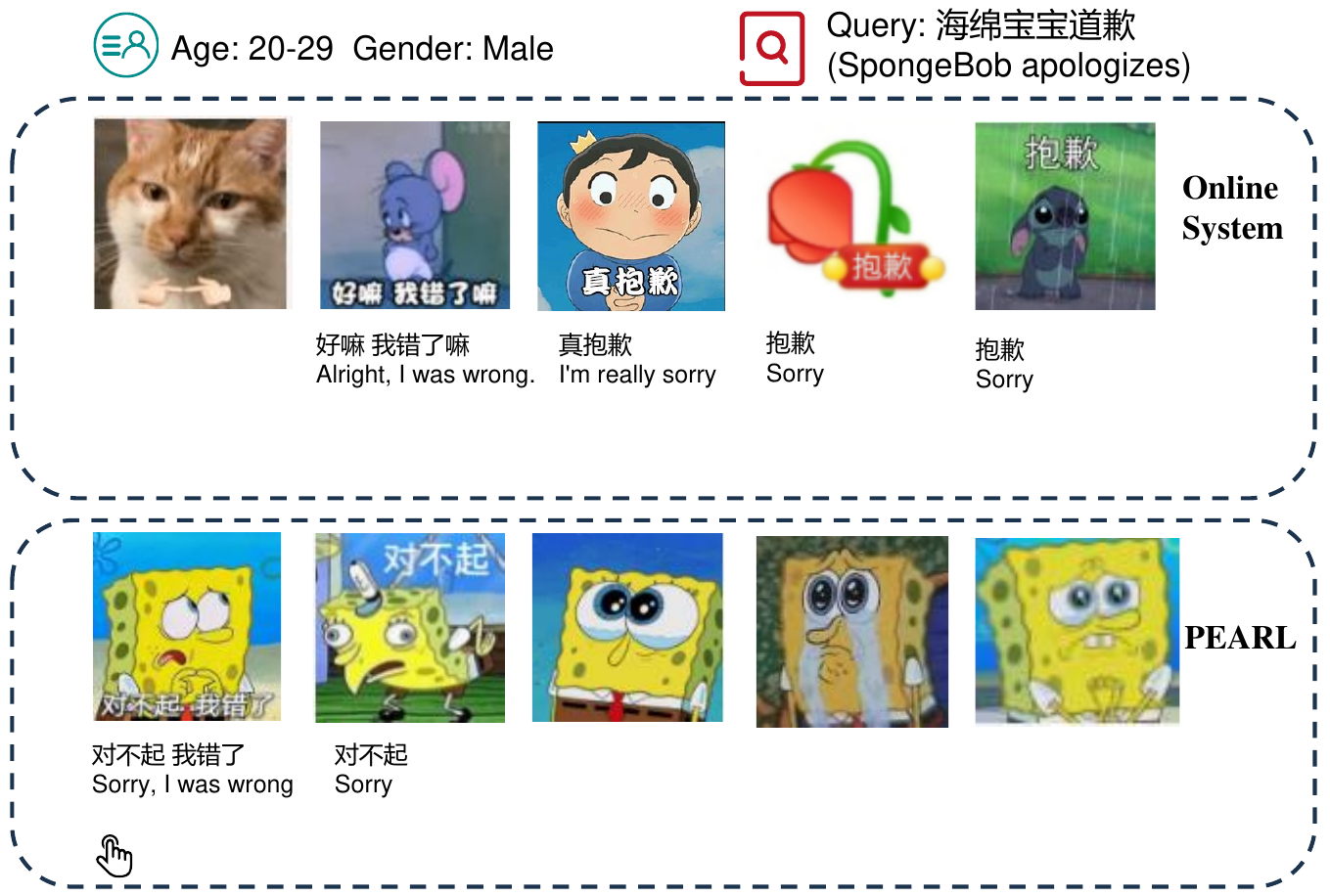}
  \vspace{-6mm} 
  \caption{Case study for the user query ``SpongeBob apologizes'' by male users aged 20-29.}
  \label{extra2}
     \vspace{-3mm}
\end{figure}

\section{Case study} \label{more case}
\autoref{extra1} and \autoref{extra2} provide two additional cases to further illustrate the advantage of PEARL.

\section{Analysis on property identifiers}\label{analysis_identifier}
We additionally conduct an analysis experiment that involves four categories of identifiers following DSI~\cite{tay2022transformer} and TIGER~\cite{rajput2023recommender}, specifically including: (i) atomic identifiers that assign an arbitrary unique integer identifier to each distinct property, (ii) string identifiers that directly utilize the property content itself as the identifier, (iii) RQ-VAE identifiers~\cite{rajput2023recommender} which utilize multi-level vector quantizer that applies quantization on residuals to generate a tuple of codewords, and (iv) PQ identifiers that yield semantic-based property identifiers through Product Quantization (PQ). The empirical results are reported in \autoref{identifier}.

According to the comparison on property identifiers, we can induce insightful findings as follows: (i) Atomic identifiers almost fail to retrieve proper stickers, which is probably due to the lack of the semantic information in unique integers. (ii) Compared to string identifiers that directly generate the property itself, RQ-VAE and PQ exhibit significantly superior retrieval effectiveness. We attribute the phenomenon to the fact that the identifier prefix tree of string identifiers is relatively less balanced, since the length of properties across distinct stickers varies a lot. In contrast, the identifiers of RQ-VAE and PQ are of the same length by applying quantization to semantic vectors, leading to a more balanced identifier prefix tree. (iii) Compared to the widely used RQ-VAE proposed by \citet{rajput2023recommender}, PQ even shows a slight advantage in terms of retrieval effectiveness. The underlying reason might be that the residual quantization process is more applicable to contents with a stronger hierarchical structure, e.g., the product metadata used in TIGER \cite{rajput2023recommender}. Nevertheless, the property content of stickers is relatively less hierarchical.

\begin{table}[h] 
\begin{tabular*}{\linewidth}{@{\extracolsep{\fill}}lcc}
\toprule
\textbf{Identifier} & \textbf{MRR@10} & \textbf{Recall@10} \\ 
\midrule
Atomic &0.0085 &0.0013 \\
String &0.0684 &0.0179 \\ 
RQ-VAE &0.1892 &0.0601 \\
PQ (ours) &\textbf{0.2074}	&\textbf{0.0732} \\
 \bottomrule
\end{tabular*}
\vspace{-2mm}
\caption{Analysis on property identifiers.} \label{identifier}
\vspace{-2mm}
\end{table}

\section{Experimental details}\label{detail}
We leverage the pyserini library~\cite{lin2021pyserini} for the implementation of BM25, DPR, and ANCE.
We leverage the fairseq library~\cite{ott2019fairseq} for the training of MINDER and PEARL.
We use the transformers library~\cite{wolf2020transformers} 
for the training of the remaining baselines, following the setup of the original literature.
All models are trained with the AdamW \cite{loshchilov2017decoupled} optimizer.
We train PEARL with a batch size of 8192 tokens and a learning rate of 1e-5.
We repeat our experiment 3 times to get the average results.
To improve efficiency, we collected the top 10,000 most frequent queries from the online system for intent analysis and precomputed their corresponding intent lists offline. During the inference time of PEARL, if a user’s query matches an entry in the offline table, the system retrieves the intent list directly without utilizing LLMs.

As for the evaluation of online tests, the full text of the instructions for the sticker-level assessment is as follows:
\texttt{Enter a query and click your favorite sticker based on your preference}.
The full text of the instructions for the session-level assessment is as follows:
\texttt{Enter a query and determine which exposure session you prefer, with the response limited to ``preferring the left exposure session'', ``preferring the right exposure session'', and ``preferring both equally''}.

\end{document}